\def\pT{\mbox{$p_T$}}
\def\sqrtsNN{\mbox{$\sqrt{s_\mathrm{NN}}$}}
\def\lt{\mbox{$<$}}
\def\gt{\mbox{$>$}}

\def\KT{\mbox{$K_{T}$}}

\documentclass[12pt]{iopart}

\usepackage{graphicx}
\usepackage{iopams}

\begin{document}

\title[High \pT\ hadron spectra at RHIC: an overview]{High \pT\ hadron spectra at RHIC: an overview}

\author{Jennifer L. Klay}
\address{Lawrence Livermore National Laboratory, Livermore, California, U.S.A. 94550}

\ead{klay@llnl.gov}

\begin{abstract}
Recent results on high transverse momentum (\pT) hadron production in p+p, d+Au and Au+Au collisions at the Relativistic 
Heavy Ion Collider (RHIC) are reviewed.  Comparison of the nuclear modification factors, $R_{dAu}(\pT)$ and $R_{AA}(\pT)$,
demonstrates that the large suppression in central Au+Au collisions is due to strong final-state effects.  Theoretical 
models which incorporate jet quenching via gluon Bremsstrahlung in the dense partonic medium that is expected in central 
Au+Au collisions at ultra-relativistic energies are shown to reproduce the shape and magnitude of the observed suppression 
over the range of collision energies so far studied at RHIC. 

\end{abstract}


\section{Introduction}
There are now four years of beam data on high \pT\ hadron spectra available from the relativistic heavy
ion collider (RHIC) covering multiple configurations (p+p, d+Au, Au+Au)
and beam energies (62.4, 130, 200 GeV).  The four RHIC experiments, 
BRAHMS, PHENIX, PHOBOS and STAR have observed and reported on several striking 
features from this new era of collider data, where the high cross-section for pQCD
hard-scattering processes plays an important role in heavy ion collisions for the 
first time.  In particular, jets of high transverse momentum hadrons which come from the
fragmentation of hard-scattered partons can be experimentally observed in heavy ion
collisions using two-particle azimuthal correlations\cite{Adler:2002ct,Adler:2002tq}.  

The yield of single inclusive high \pT\ charged hadrons leading these jets is observed to be substantially
suppressed (by a factor of 3-5 for \pT\gt 6 GeV/c) in the most central Au+Au collisions
at \sqrtsNN = 130 and 200 GeV compared to the expectation from scaling the yield in p+p
collisions at the same beam energy by the average number of binary nucleon-nucleon collisions
in Au+Au\cite{Adcox:2001jp,Adler:2002xw+Adams:2004zg,Adams:2003kv,Adler:2003au}.  The yields in d+Au 
collisions are slightly enhanced at intermediate \pT\ and consistent with binary-collision scaling for 
\pT$\gtrsim$7 GeV/c\cite{Back:2003ns,Adler:2003ii,Adams:2003im,Arsene:2003yk}.  

Detailed studies of the two-particle azimuthal correlations as a function of centrality in
Au+Au collisions at \sqrtsNN = 200 GeV reveal that the correlations at $d\phi=\pi$ expected
from the balancing partner jet created in the partonic hard-scattering become more and more
suppressed in central Au+Au collisions in stark contrast to p+p and d+Au collisions.  The back-to-back
partner of the di-jet effectively ``disappears'' into the bulk matter soft hadronic background
generated in the collision\cite{Adler:2002tq,Adams:2003im}.

These features, along with the observed large azimuthal anisotropy of hadron distributions at large \pT\ in 
non-central Au+Au collisions\cite{Adler:2002ct,Adams:2004wz}, provide strong evidence that a dense, strongly 
interacting medium is formed in Au+Au collisions at RHIC, substantially effecting the production and 
dynamics of jets which traverse it.  In this article I review the available data on the transverse momentum 
spectra of leading hadrons in p+p, d+Au and Au+Au collisions and how the theoretical interpretations of 
these data contribute to our understanding of the dense nuclear medium created in central heavy ion 
collisions at RHIC.

\section{The baseline:  high \pT\ particle spectra in elementary hadron collisions}

\begin{figure}
\centering
\includegraphics[width=.4\textwidth]{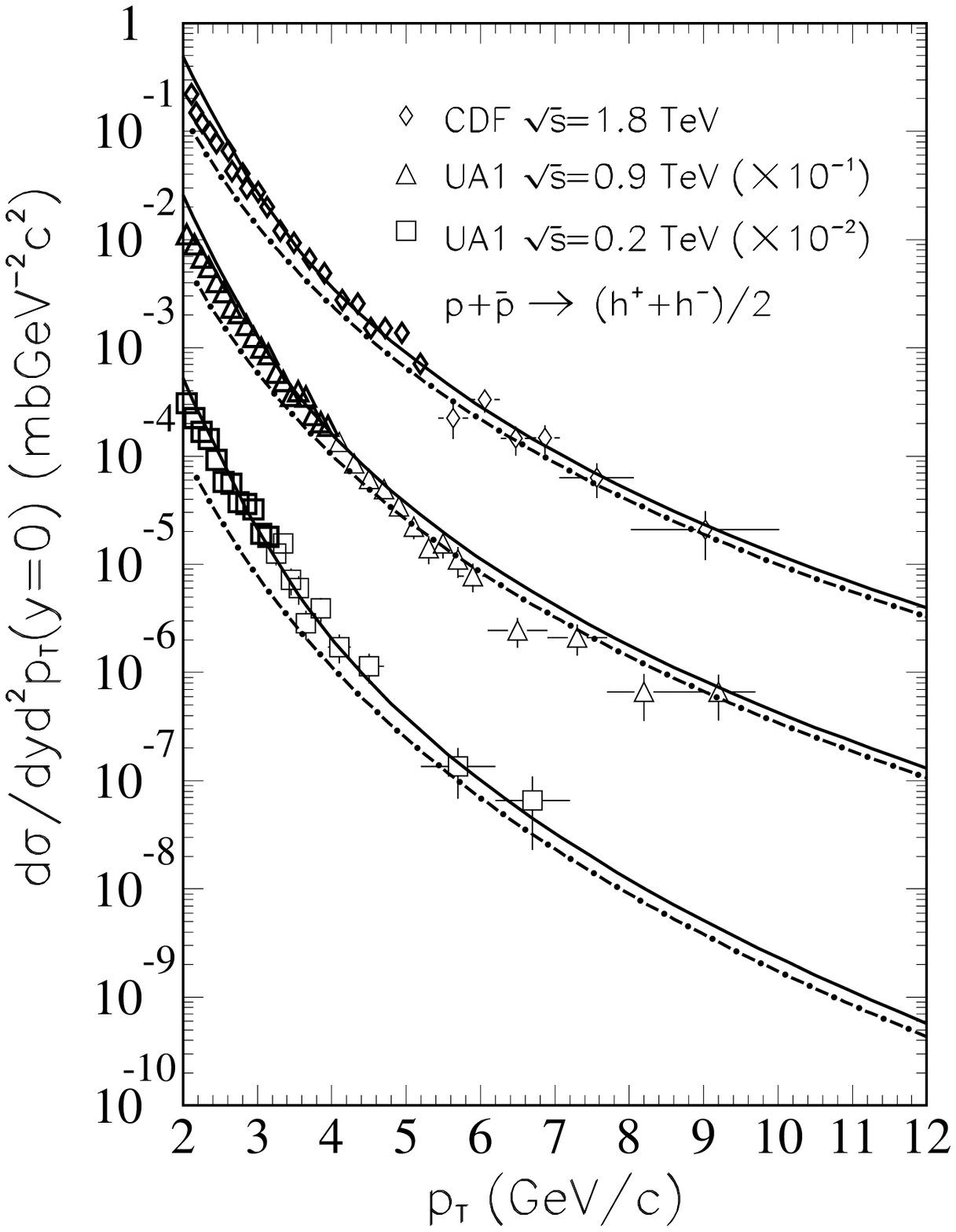}
\hspace{.1\textwidth}
\includegraphics[width=.45\textwidth]{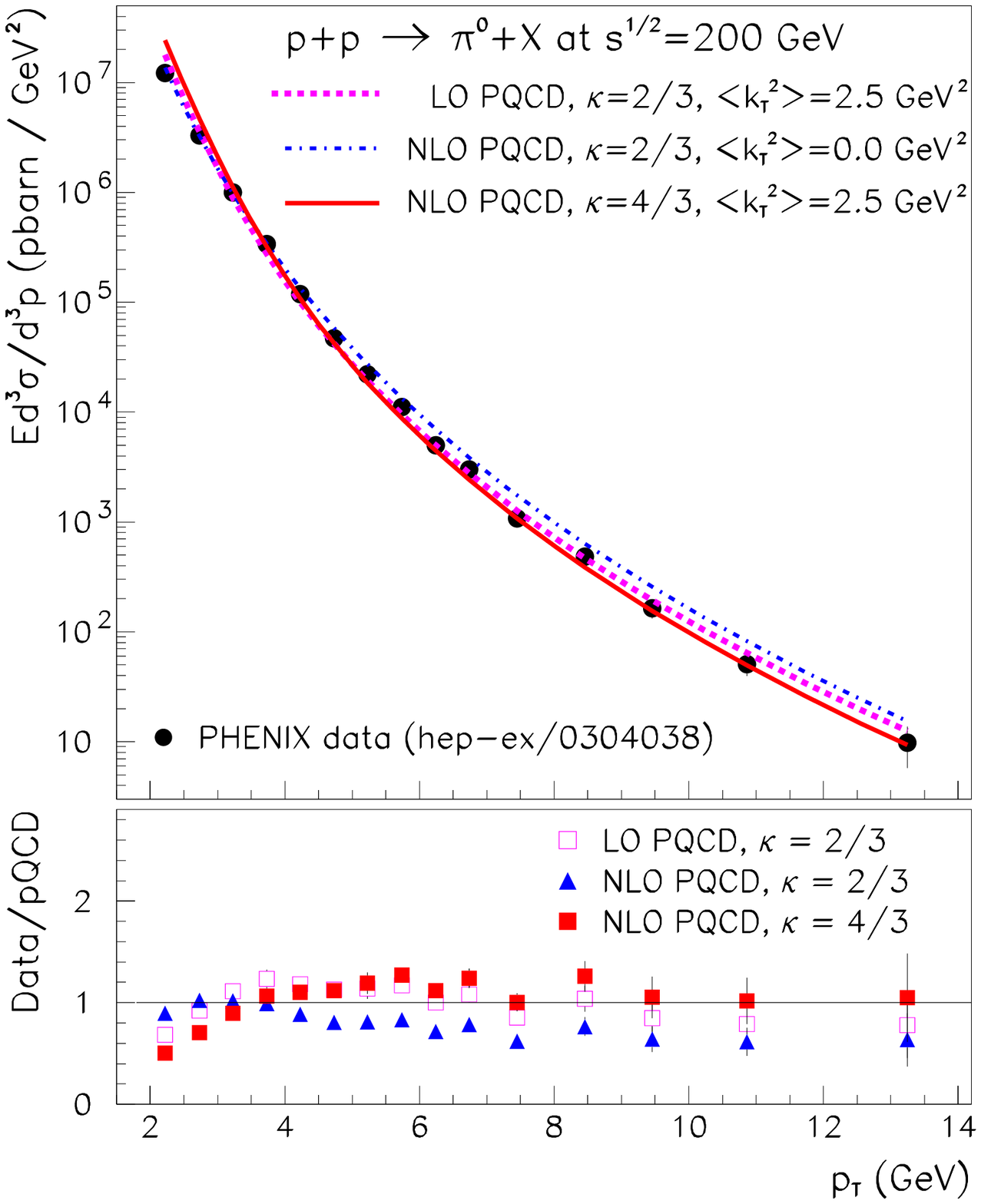}
\caption{\label{Fig1} Comparisons of $h^{\pm}$ and $\pi^{0}$ spectra from p+p($\bar{p}$) collisions to leading order and 
next-to-leading order pQCD parton model calculations.  The left panel is from \cite{Wang:1998ww} and the right panel is 
from \cite{Levai:2003at}.}
\end{figure}

Within the perturbative QCD (pQCD) parton model, the short-distance perturbatively calculable hard-scattering 
cross-section factorizes from the long-distance non-perturbative parton distribution functions 
(PDFs) and parton fragmentation functions, providing a way to describe high \pT\ particle production in 
elementary collisions\cite{Collins:1985ue}.  The parton distribution functions are determined 
from deep inelastic scattering data, while the fragmentation functions are obtained 
from experimental studies of e+e- collisions.  Within the pQCD parton model, the expression 
for the production of high transverse momentum hadrons in pp($\bar{p}$) collisions can be written

\begin{equation}
  \frac{d\sigma^h_{pp}}{dyd^2p_T} = K \sum_{abcd}
  \int dx_a dx_b f_a(x_a,Q^2)f_b(x_b,Q^2)
  \frac{D^0_{h/c}(z_c,Q^2)}{\pi z_c}
  \frac{d\sigma^{(ab\rightarrow cd)}}{d\hat{t}}, 
\label{eq:nch_pp}
\end{equation}

\noindent where $f_a(x_a,Q^2)$ and $f_b(x_b,Q^2)$ represent the distribution of partons 
carrying momentum fraction $x$ within the incoming hadrons, $\frac{D^0_{h/c}(z_c,Q^2)}{\pi z_c}$ is the 
fragmentation function to convert parton $c$ into colorless hadrons, $h$ and 
$\frac{d\sigma^{(ab\rightarrow cd)}}{d\hat{t}}$ is the hard-scattering cross-section, which is calculable in 
a perturbative series expansion in powers of the strong coupling constant, $\alpha_{s}$.  Since the 
complexity of the calculation grows nonlinearly with the power of $\alpha_{s}$, leading order calculations 
with a phenomenological K-factor to account for the next-to-leading order corrections are 
often used.  For \sqrtsNN$\gtrsim$50 GeV, these calculations agree impressively well 
with the experimental data.  In the left panel of Figure \ref{Fig1}, a parton model calculation is 
compared to data from pp($\bar{p}$) collisions at three collision energies\cite{Wang:1998ww}.  The 
right panel of Figure \ref{Fig1} shows a comparison of both LO and NLO calculations to $\pi^{0}$ production 
in p+p collisions at RHIC\cite{Levai:2003at}.  The addition of terms to account for finite transverse 
momentum \KT\ of the partons within the nucleus improves the agreement with the experimental data in both 
cases.  The success of this description in describing the hadron spectra in p+p($\bar{p}$) collisions 
suggests that a similar formalism may be a useful starting point for the more complex systems: p(d)+A and 
A+A.

\section{Leading particle spectra in heavy ion collisions}

In p(d)+A and A+A collisions, additional nuclear medium effects must be taken into account.
The parton distribution functions in nuclei differ from those in free nucleons\cite{Emel'yanov:1999bn}.  The 
distributions of partons with $x \lesssim 10^{-1}$ are suppressed relative to the same distributions in 
nucleons, a phenomenon known as ``shadowing''.  For partons with $x\sim$ 1 the distributions are 
enhanced or ``anti-shadowed''.  In addition, multiple soft scattering of the projectile partons as they 
traverse a target nucleus may boost their transverse momentum slightly even before they undergo the 
hard-scattering or subsequent to it, leading to an enhancement of the yield at high \pT\ compared to p+p 
collisions\cite{Lev:1983hh+Ochiai:1985ie}.  These effects can be added to the factorized pQCD model expressed 
in Equation \ref{eq:nch_pp} as modifications to the PDF terms.

\begin{figure}
\centering
\includegraphics[width=.45\textwidth]{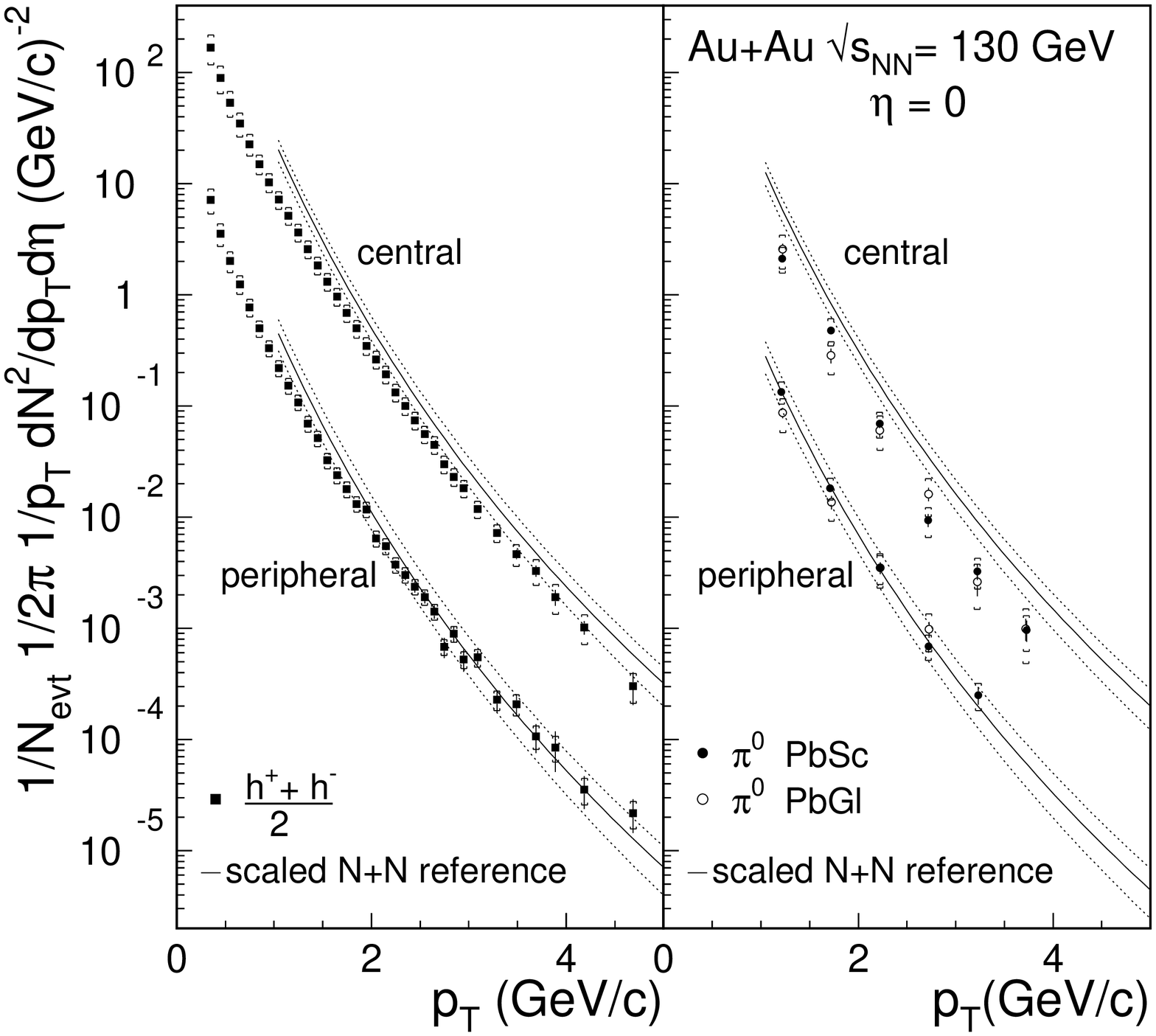}
\hspace{.05\textwidth}
\includegraphics[width=.45\textwidth]{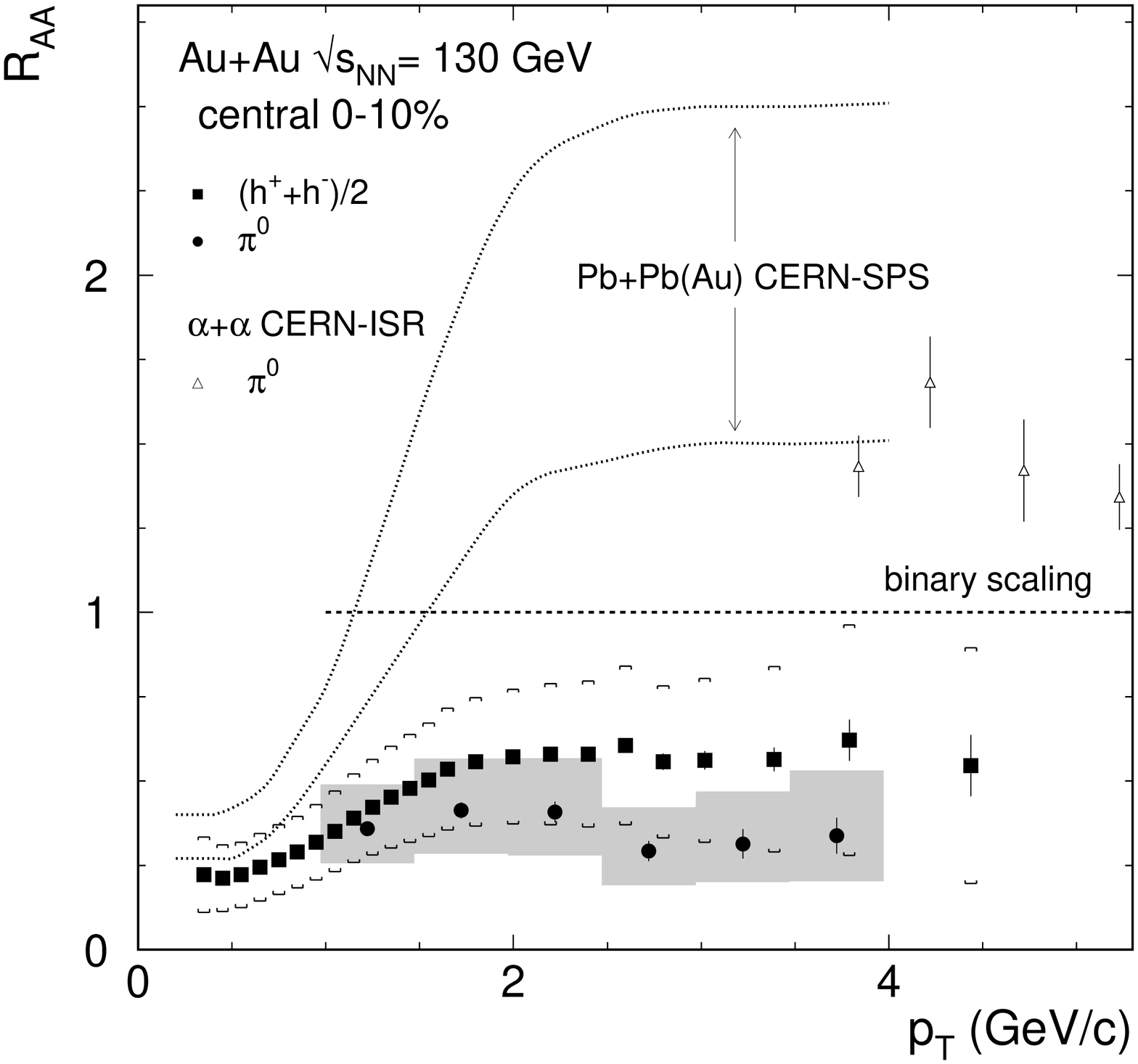}
\caption{\label{Fig2} Transverse momentum spectra and nuclear modification factor $R_{AA}$ of $h^{\pm}$ and 
$\pi^{0}$ measured by the PHENIX experiment for Au+Au collisions at \sqrtsNN = 130 GeV\cite{Adcox:2001jp}.}
\end{figure}

Finally, the nuclear medium produced in a collision might influence the production of high 
\pT\ hadrons, either through hard partonic or hadronic re-interactions in the final-state after the 
hard-scattering.  Partonic interactions with a colored medium induce gluon Bremsstrahlung radiation, 
reducing the parton's energy before it fragments into hadrons\cite{Baier:2000mf+Gyulassy:2003mc}.  
This effect is often introduced in pQCD parton model calculations through a modified fragmentation 
function\cite{Wang:1996pe+Wang:1996yh,Wang:1998bh} or in the form of quenching weights\cite{Salgado:2003gb}, for example.   

By comparing the yields in p(d)+A and A+A collisions to that in p+p collisions, with a scaling 
factor to take into account the nuclear geometry, one can test the assumption that a 
nucleus-nucleus collision is a simple superposition of incoherent nucleon-nucleon scatterings and explore 
these nuclear effects.  The magnitude and shape of deviations from this scaling in Au+Au collisions may be 
used to infer information about the properties of the medium\cite{Wang:1998bh,Vitev:2002pf}.  This is the 
fundamental goal of studying high \pT\ particle production and dynamics in heavy ion collisions:  What can 
be learned about the medium from these ``hard probes''?

The nuclear modification factor, $R_{AB}$, is defined as

\begin{equation}
  R_{AB}(p_T,b)\equiv \frac{d\sigma^h_{AB}/dyd^2p_T(b)}
  {\langle N_{\rm binary}(b)\rangle d\sigma^h_{pp}/dyd^2p_T} \label{eq:ratio}.
\end{equation}

\noindent where $d\sigma^h_{AB}/dyd^2p_T(b)$ is the differential cross-section for hadron production in a nuclear 
collision of impact parameter, b, $d\sigma^h_{pp}/dyd^2p_T$ is the corresponding quantity in p+p collisions 
and $\langle N_{\rm binary}(b)\rangle$ is the average number of binary nucleon-nucleon collisions, often 
calculated using the Glauber model for the nuclear overlap with given impact parameter\cite{Glauber:1959}.

\begin{figure}
\includegraphics[width=.45\textwidth]{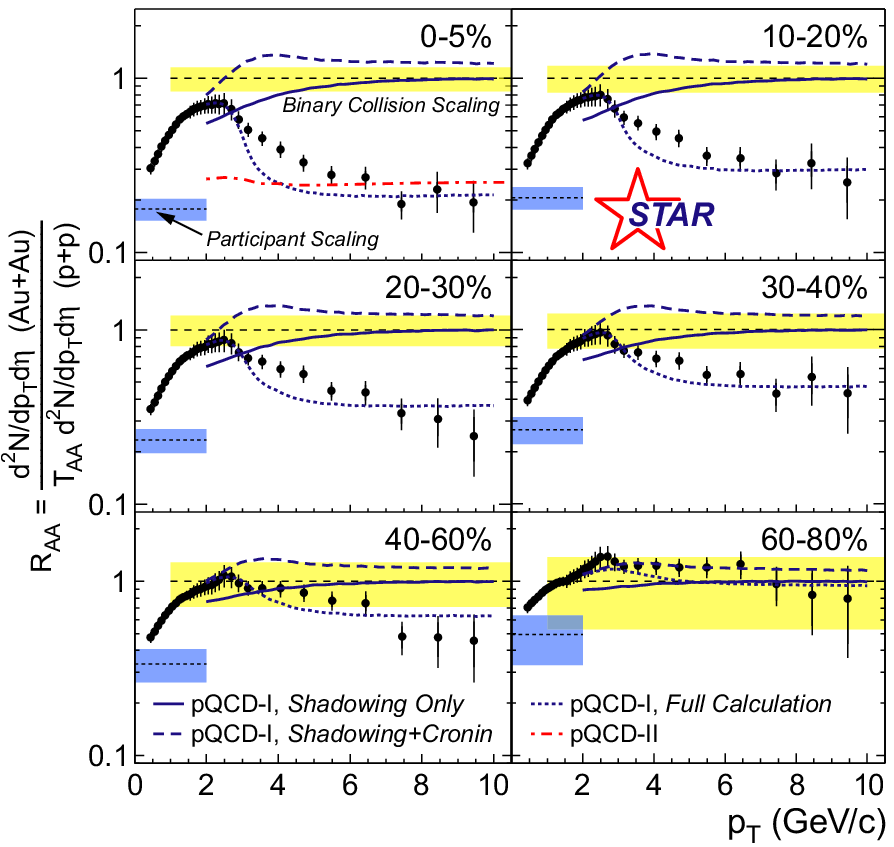}
\hspace{.05\textwidth}
\includegraphics[width=.45\textwidth]{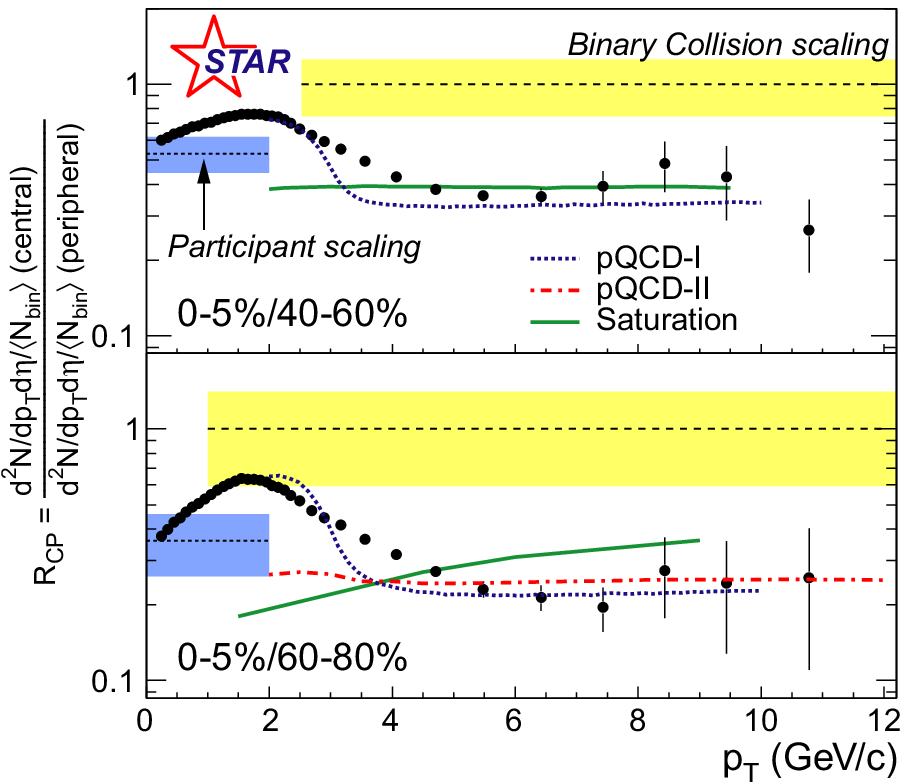}
\caption{\label{Fig3} Nuclear modification factors, $R_{AA}$ (left) and $R_{CP}$ (right) for $h^{\pm}$ measured by the 
STAR experiment in Au+Au collisions at \sqrtsNN = 200 GeV\cite{Adams:2003kv}.}
\end{figure}

In the first collision data from RHIC at \sqrtsNN=130 GeV, the PHENIX experiment showed that 
the production of high \pT\ $h^{\pm}$ and $\pi^{0}$'s was suppressed by a factor of $\sim$3 in 
central Au+Au collisions compared to the expectation from binary-scaled p+p collisions, as 
seen in Figure \ref{Fig2} \cite{Adcox:2001jp}.  This stunning result, also seen in $h^{\pm}$ spectra by
the STAR experiment\cite{Adler:2002xw+Adams:2004zg}, confirmed that a qualitatively new kind of system with intriguing 
properties is generated in these collisions.  However these first data are {\em quantitatively} limited by 
the parameterized p+p reference spectrum used for the comparison.  The reference had to be derived from a 
compilation of the world's data on hadron production in p+p($\bar{p}$) collisions over a broad range of 
collision energies because no data exist at \sqrtsNN = 130 GeV.  

In RHIC Run II, this limitation was overcome when data from p+p and Au+Au
collisions at \sqrtsNN = 200 GeV were collected\cite{Adams:2003kv,Adler:2003qi,Back:2003qr,Arsene:2003yk}.  
Figure \ref{Fig3} shows results for $h^{\pm}$ $R_{AA}$ on the left and a similar measure, $R_{CP}$, 
the binary collision-scaled yield in central Au+Au compared to the scaled yield in peripheral Au+Au 
collisions\cite{Adams:2003kv} on the right.  This comparison, which is motivated by the observed binary 
collision-scaling in peripheral Au+Au collisions for \pT\gt 2 GeV/c seen in the lower panels of the left 
side of Figure \ref{Fig3}, extends the \pT\ reach of the measurement with smaller systematic uncertainties 
than $R_{AA}$.  The high statistics data sample at \sqrtsNN = 200 GeV, with smaller uncertainties and higher 
\pT\ reach than \sqrtsNN = 130 GeV, reveals the feature that for all hadrons with 6\lt\pT\lt12 GeV/c, the  
suppression is approximately independent of \pT.  This \pT-independence was largely unexpected in model 
predictions prior to the availability of the data\cite{Wang:1998ww}.  

\begin{figure}
\centering
\includegraphics[width=.42\textwidth]{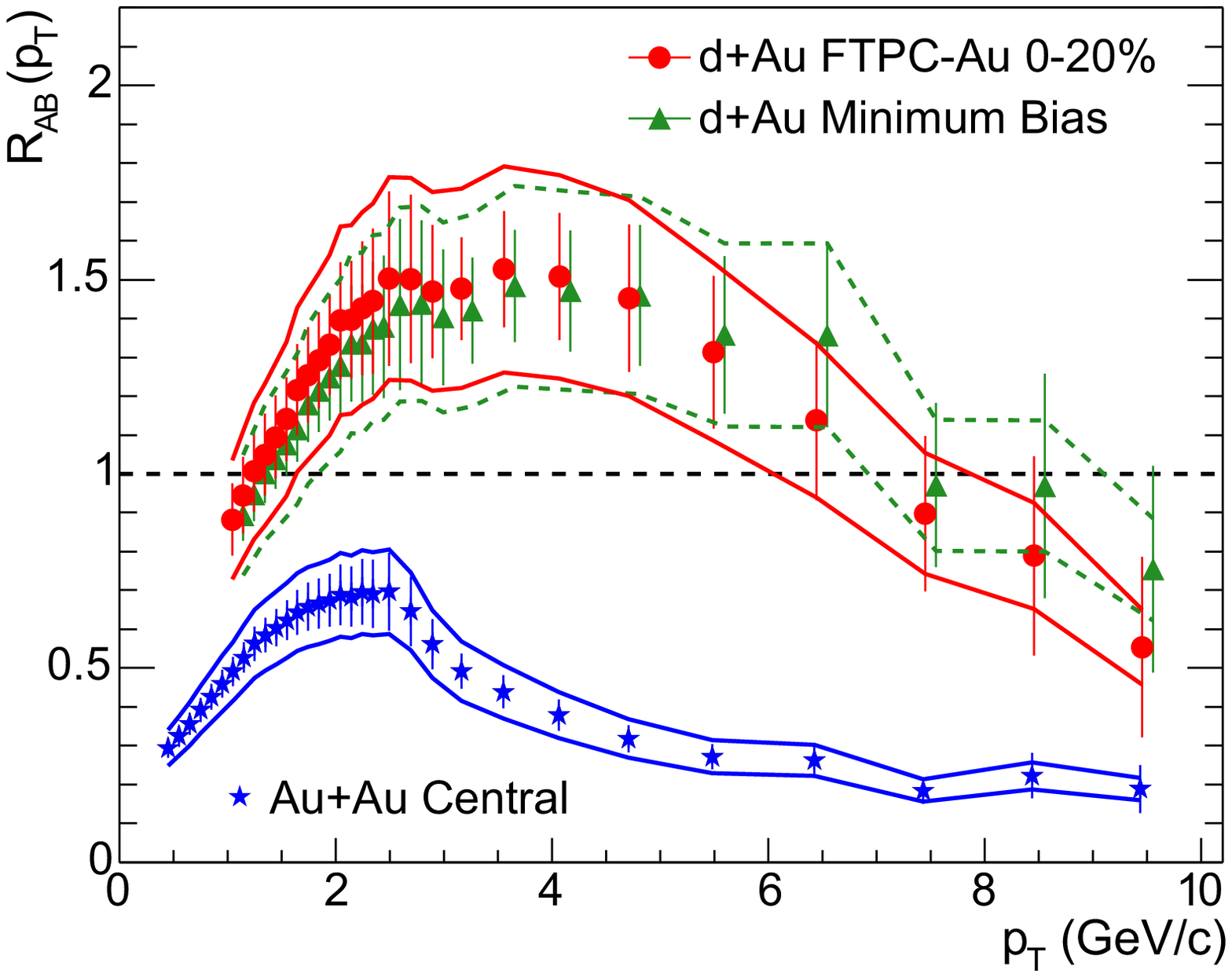}
\hspace{5mm}
\includegraphics[width=.46\textwidth]{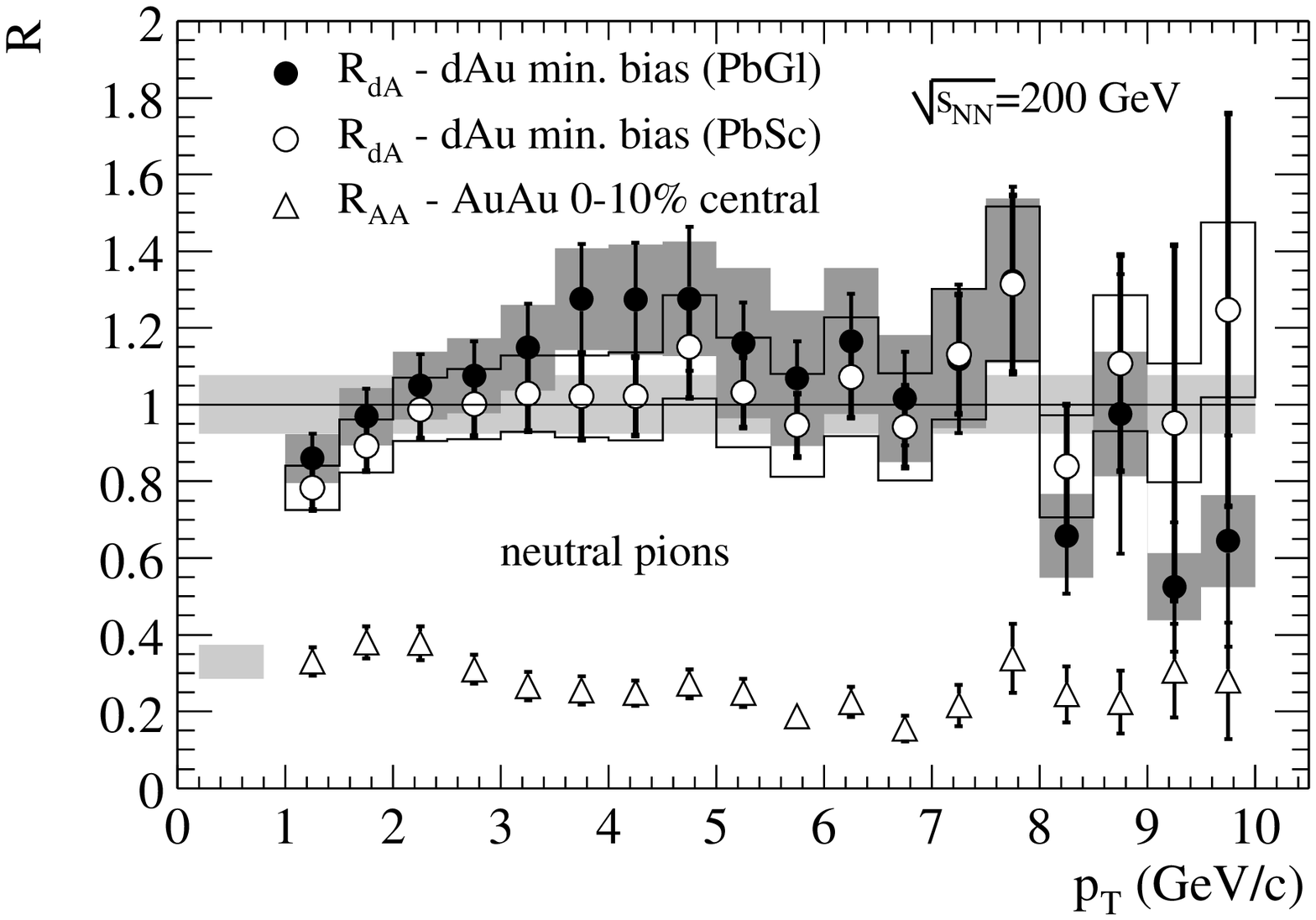}
\vspace{2mm}
\includegraphics[width=.45\textwidth]{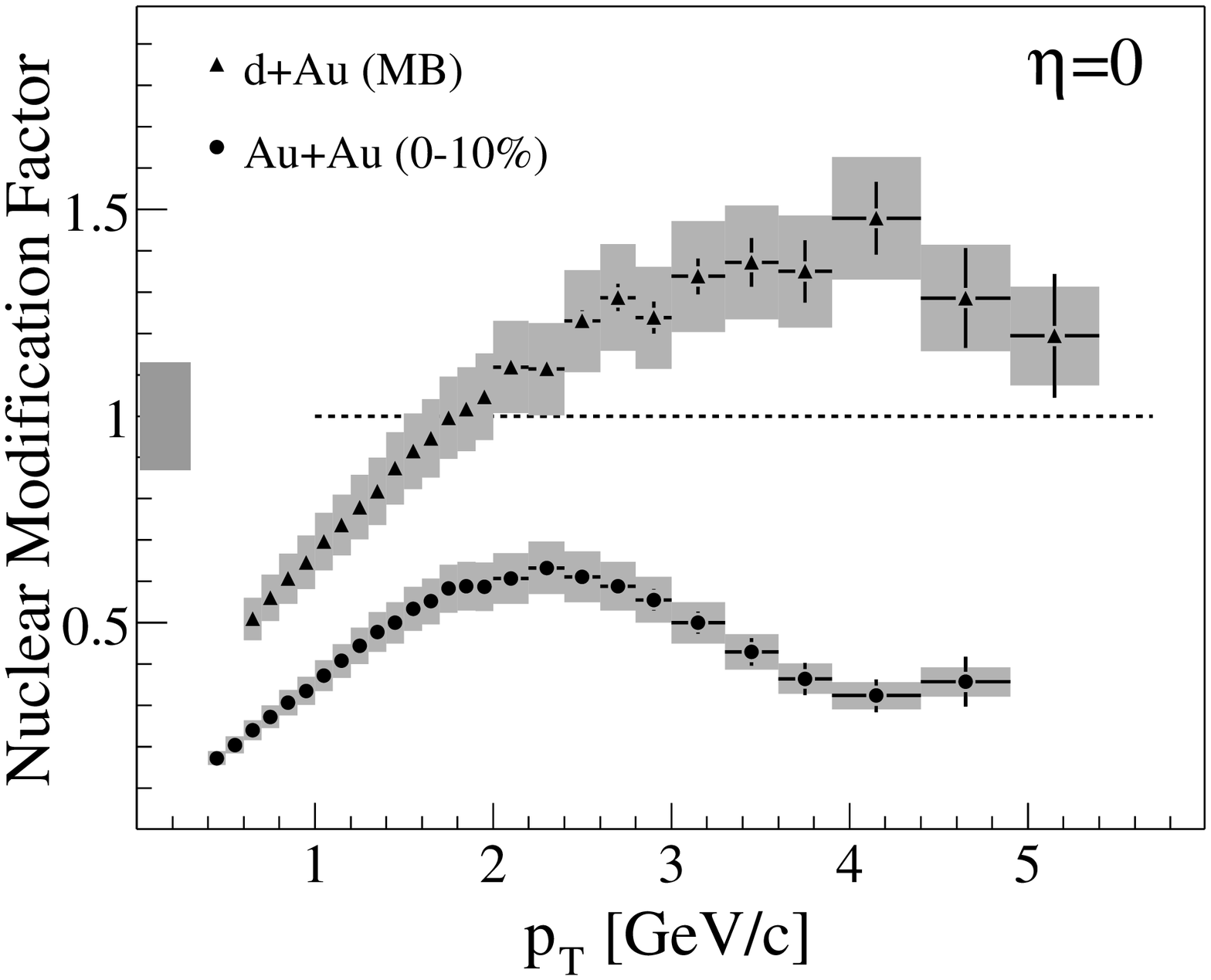}
\hspace{2mm}
\includegraphics[width=.42\textwidth]{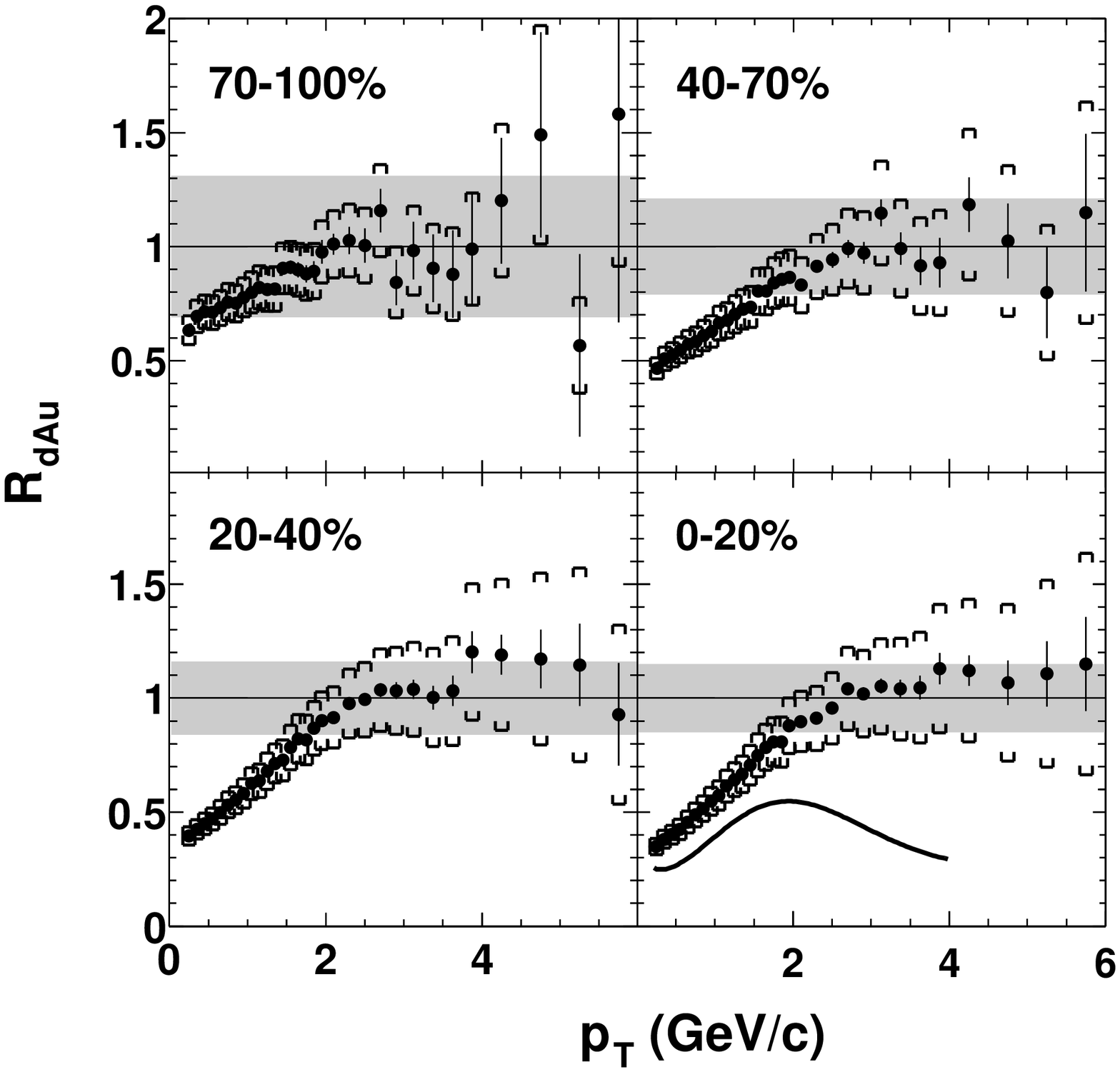}
\caption{\label{Fig4} Nuclear modification factors $R_{dAu}$ in d+Au collisions at \sqrtsNN = 200 GeV.  Clockwise from the 
upper left: STAR $h^{\pm}$\cite{Adams:2003im}, PHENIX $\pi^{0}$\cite{Adler:2003ii}, PHOBOS 
$h^{\pm}$\cite{Back:2003ns} and BRAHMS $h^{\pm}$\cite{Arsene:2003yk}.}
\end{figure}

More recent model calculations are compared to the data in Figure \ref{Fig3}.  Two formulations of a pQCD 
parton model incorporating partonic energy loss, labeled pQCD-I\cite{Wang:2003mm} and 
pQCD-II\cite{Vitev:2002pf} are shown.  The influence of nuclear gluon shadowing is shown within pQCD-I as 
the dashed lines of the left panel of Figure \ref{Fig3}.  It is the subtle interplay of the shadowing, 
initial state multiple scattering and jet quenching which results in the observed \pT-independence in these 
models.  In addition, the right panel of Figure \ref{Fig3} also shows a calculation from the 
parton saturation model which extends initial-state gluon saturation beyond the saturation scale, $Q_{s}$ to 
high \pT\ at midrapidity\cite{Kharzeev:2002pc}.  This model is also seen to reproduce the suppression pattern 
for \pT\gt 5 GeV/c in the upper panel.   The fact that two models based on entirely different physics, one 
incorporating strong initial-state effects and one with strong final-state effects, can both explain the 
suppression data was the primary reason that d+Au collisions were run at RHIC.  Initial-state nuclear 
effects are present in both d+Au and Au+Au collisions, while final-state effects are expected only in Au+Au.  

Figure \ref{Fig4} shows results on $R_{dAu}$ from each of the four RHIC 
experiments\cite{Back:2003ns,Adler:2003ii,Adams:2003im,Arsene:2003yk}.  The enhancement 
observed for 2\lt\pT\lt6 GeV/c is similar to the enhancement seen in p+A collisions at lower 
energies\cite{Cronin:1974zm+Antreasyan:1979cw}, called the ``Cronin effect''.  The standard explanation of 
this observed enhancement is the multiple soft interactions prior to and possibly after the hard-scattering which were discussed 
in the first section\cite{Lev:1983hh+Ochiai:1985ie}.

\begin{figure}
\centering
\includegraphics[width=.45\textwidth]{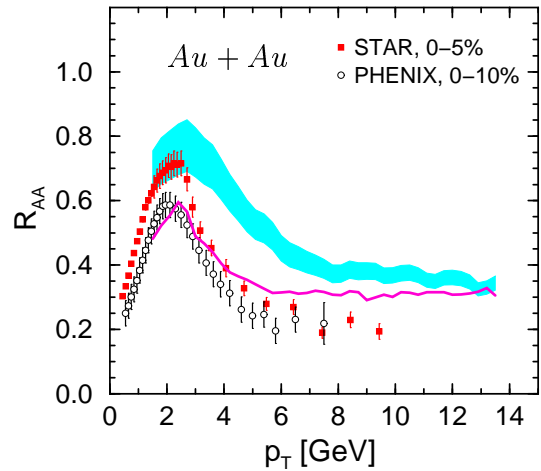}
\caption{\label{Fig5} A comparison of results from the Hadron-String-Dynamics model\cite{Cassing:2003sb} 
to RHIC data on $R_{AA}$ in Au+Au collisions at \sqrtsNN = 200 GeV. }
\end{figure}

In contrast, the gluon saturation model shown in the right panel of Figure \ref{Fig3} predicted a 
suppression for central d+Au collisions of $R_{dA} \simeq$0.7\cite{Kharzeev:2002pc}.  Clearly no such 
suppression is present in Figure \ref{Fig4}.  The factor 4-5 suppression in central Au+Au 
collisions must be attributed to final-state, not initial-state nuclear effects.  

Although the data in Figure \ref{Fig3} compare quite well with models incorporating 
final-state partonic energy loss, final state hadronic re-scattering may play some role in 
generating the observed suppression and many theoretical investigations of this possibility 
have been undertaken\cite{Gallmeister:2002us,Lietava:2003df,Capella:2004xj}.  In Figure \ref{Fig5}, a comparison is made 
between RHIC data and a Hadron-String-Dynamics (HSD) model which incorporates quark, diquark, string and hadronic 
interactions with the medium\cite{Cassing:2003sb}.  The features of the data are indeed reproduced, but hadronic 
interactions are found to contribute very little to the suppression for \pT\gt 6 GeV/c, where inelastic, possibly colored 
partonic interactions of leading ``pre-hadrons'' with the dense medium are responsible for the observed 
suppression.  

\section{Identified particle spectra and pseudorapidity dependence of $R_{AA}$}

\begin{figure}
\centering
\includegraphics[width=.45\textwidth]{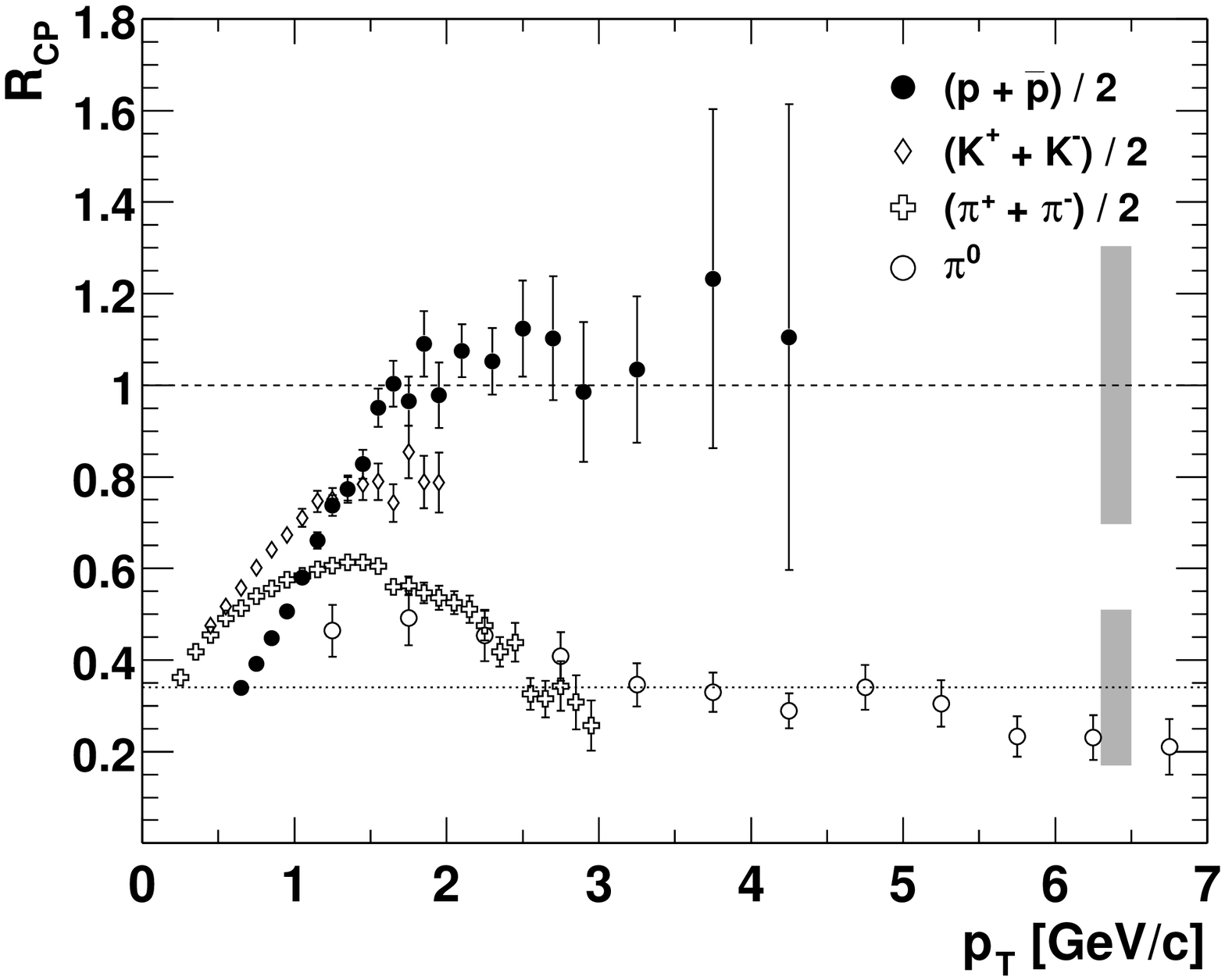}
\hspace{5mm}
\includegraphics[width=.45\textwidth]{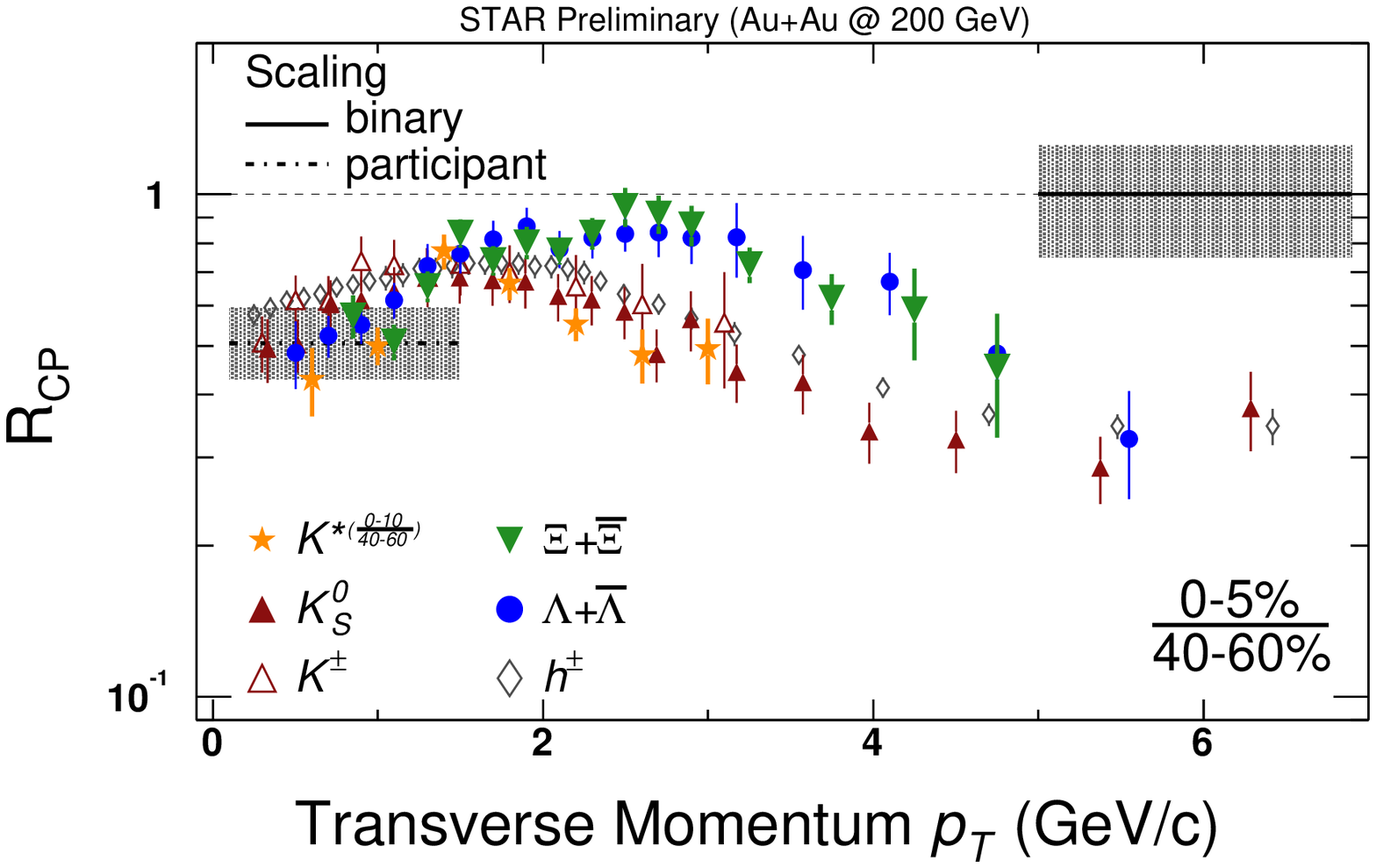}
\caption{\label{Fig6} Nuclear modification factor $R_{CP}$ for identified hadrons from PHENIX\cite{Adler:2003au} on the 
left and STAR\cite{Lamont:2004qy} on the right.}
\end{figure}

The particle composition at intermediate \pT\ has been observed to contain an unusually 
bountiful baryon population in Au+Au collisions at RHIC\cite{Adler:2003kg,Adams:2003am}.  Figure \ref{Fig6} 
shows results from PHENIX\cite{Adler:2003au} on the left and STAR\cite{Lamont:2004qy} on the right for 
$R_{CP}$ of identified hadrons.  The difference between meson and baryon $R_{CP}$ is most pronounced for 
2\lt\pT\lt6 GeV/c.  Several models relying on parton recombination and coalescence have been proposed to 
explain the ``baryon anomaly'' at intermediate \pT\cite{Hwa:2003bn+Fries:2003vb+Greco:2003xt}.  From the 
right panel of Figure \ref{Fig6}, it appears that the mesons and baryons converge for \pT\gt6 GeV/c, where 
parton fragmentation dynamics dominate and no difference between mesons and baryons is expected.  Interesting 
dynamics are clearly at play in the intermediate \pT\ region, which should be investigated more thoroughly.   
What is also clear is the importance of carrying all leading hadron observables, such as spectra, 
di-hadron correlations and azimuthal anisotropies to the highest \pT\ available and at minimum for \pT\gt 6 
GeV/c.  This ensures that they can be used as reliable hard probes of the medium properties while minimizing 
complications from the dynamics of additional effects.

\begin{figure}
\centering
\includegraphics[width=.5\textwidth]{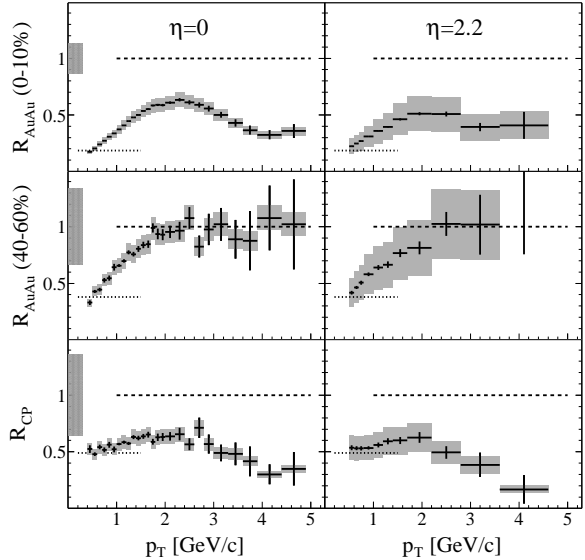}
\caption{\label{Fig7} Nuclear modification factors $R_{AA}$ and $R_{CP}$ in Au+Au collisions at \sqrtsNN = 200 GeV for 
midrapidity (right) and forward rapidity (left) measured by the BRAHMS collaboration\cite{Arsene:2003yk}.}
\end{figure}

\begin{figure}
\centering
\includegraphics[width=0.9\textwidth]{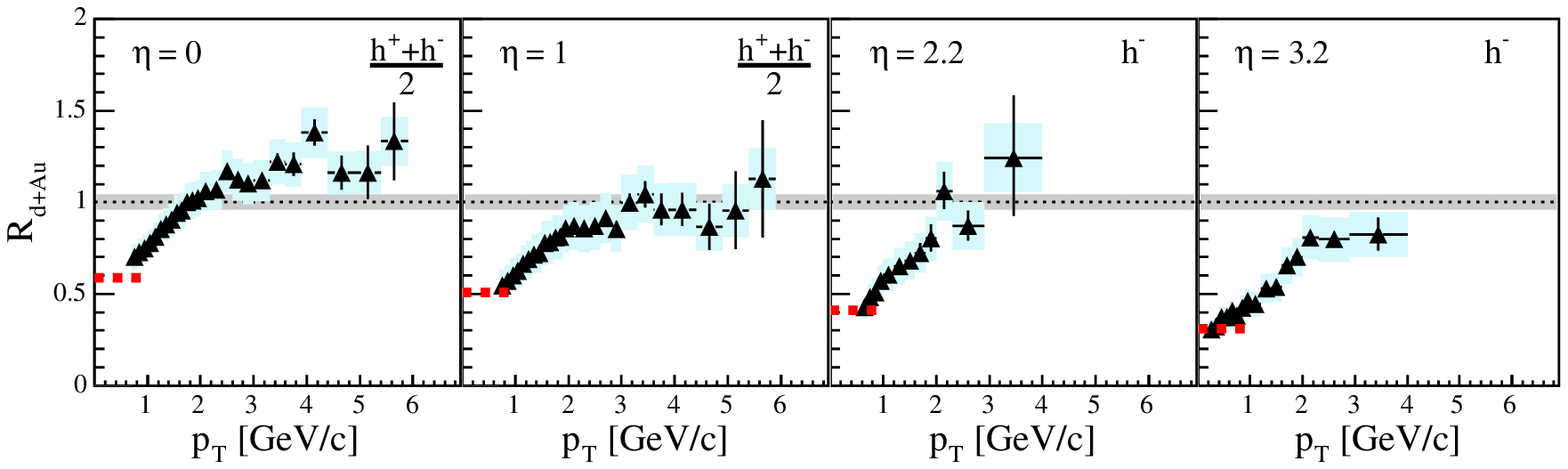}
\vspace{2mm}
\includegraphics[width=0.9\textwidth]{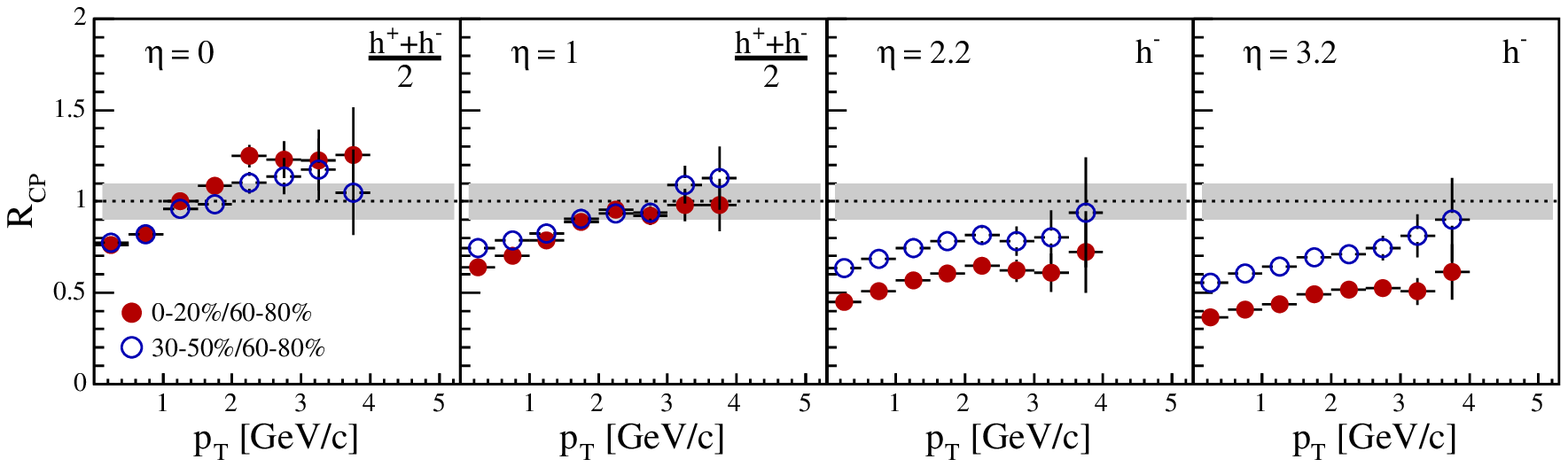}
\caption{\label{Fig8} Nuclear modification factors $R_{dAu}$ (top) and $R_{CP}^{dAu}$ (bottom) for four rapidity 
intervals in d+Au collisions at \sqrtsNN = 200 GeV measured by the BRAHMS Collaboration\cite{Arsene:2004ux}.}
\end{figure}

In Figures \ref{Fig7} and \ref{Fig8}, data from the BRAHMS collaboration on $R_{AA}$ and 
$R_{CP}$ at forward rapidities are compared to the results at mid-rapidity in Au+Au\cite{Arsene:2003yk} and 
d+Au\cite{Arsene:2004ux} collisions.  From Figure \ref{Fig7} it appears that there is no strong evolution of the 
suppression pattern as a function of pseudorapidity in Au+Au, while in Figure \ref{Fig8} there is a clear 
change from Cronin enhancement at midrapidity to a suppression at forward rapidities in the Au-going 
direction of d+Au collisions.  Such a pattern is expected in saturation 
models\cite{Baier:2003hr+Albacete:2003iq,Kharzeev:2003wz,Kharzeev:2002pc,Jalilian-Marian:2003mf+Dumitru:2002qt} 
but may not be a unique signature of gluon saturation.  In fact, energy-momentum conservation in the hadron 
fragmentation region alone should result in a reduction of $R_{dAu}$ of about 30\% between
$\eta = 0$ and $\eta = 3.2$\cite{Capella:2004xj}.  In addition, conventional nuclear shadowing corrections 
are also expected to result in a roughly 30\% decrease of $R_{dAu}$ over the same range of 
pseudorapidity.  The authors of \cite{Capella:2004xj} find that they can reproduce the data over 
all rapidities incorporating these effects into a hadronic absorption model without invoking 
substantial gluon saturation in the initial wave-function of the nucleus.  More studies of the pseudorapidity dependence of $R_{AB}$ and other 
high \pT\ observables such as di-hadron correlations are probably needed before a concrete understanding of the underlying dynamics emerges.

\section{Evolution of $R_{AA}$ with beam energy: Enhancement vs. Suppression?}

A Bjorken model estimate of the initial energy density achieved in central Au+Au 
collisions at RHIC yields values 30 times that in cold nuclear matter\cite{Adcox:2001ry} 
while at the SPS the corresponding value is approximately a factor of 20\cite{Margetis:1994tt}.  
These values are both well above the factor 3-5 expected for the onset of deconfinement and yet the 
yield of $\pi^{0}$ measured at the SPS appears enhanced by a factor of 2-4 relative to binary collision 
scaling\cite{Aggarwal:2001gn}.  Is there a transition from enhancement to suppression somewhere 
between 17\lt\sqrtsNN\lt130 GeV?

\begin{figure}
\centering
\includegraphics[width=.45\textwidth]{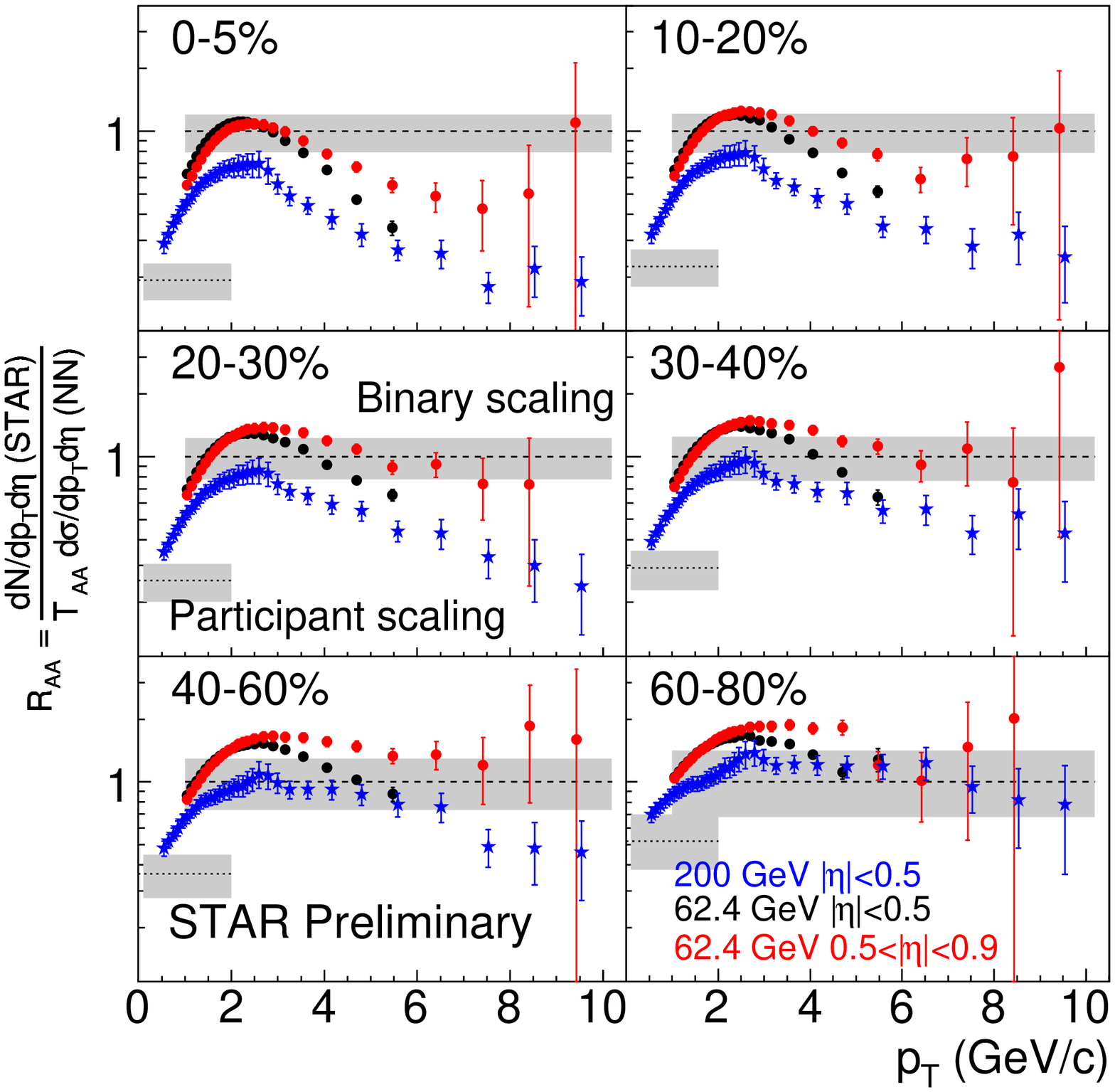}
\vspace{2mm}
\includegraphics[width=.45\textwidth]{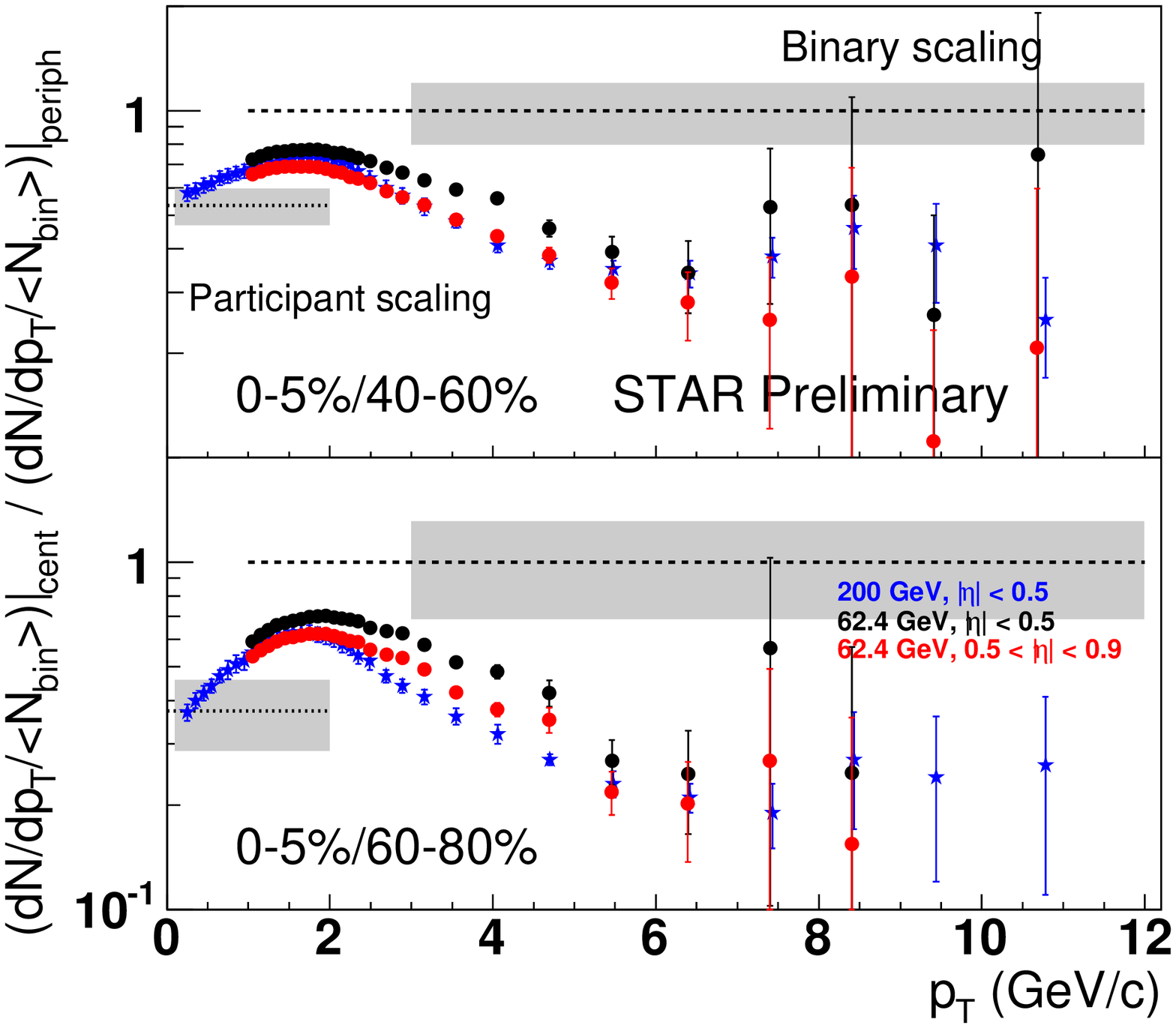}
\caption{\label{Fig9} Preliminary results from the STAR Experiment on $R_{AA}$ and $R_{CP}$ in Au+Au collisions at 
\sqrtsNN = 62.4 GeV\cite{RHICAGSUsersSTAR}.}
\end{figure}

During RHIC Run IV, Au+Au collisions at \sqrtsNN = 62.4 GeV were run to search for such a 
transition.  The choice of beam energy was motivated by the fact that it lies  
roughly halfway between the SPS and RHIC top energies on a logarithmic scale and there
exists a large body of reference data from p+p collisions at the ISR at this beam 
energy\cite{Akesson:1982sc+Breakstone:1995ug+Drijard:1982zk+Angelis:1978ec+Busser:1976su+Akesson:1984iq+Eggert:1975cq+Alper:1975jm+Alper:1974rw}.
As it turns out, two key findings underscoring the importance of a reliable reference 
dataset have emerged from this exploration.

First, a recent re-analysis of the parameterization used for the SPS $R_{AA}$ reference reveals 
large variations in the extracted $R_{AA}$ depending on the choice of data and functional 
form used to fit them\cite{d'Enterria:2004ig}.  In fact, a parameterization using the best fit to 
all of the available data results in $R_{AA}$ for $\pi^0$ with 2\lt\pT\lt4 GeV/c that is 
consistent (within large uncertainties) with binary collision scaling at \sqrtsNN=17 GeV.  
$R_{CP}$ may even be slightly suppressed at the SPS\cite{d'Enterria:2004ig}.  

Second, it is challenging to find a reference spectrum within the ISR 
datasets\cite{Akesson:1982sc+Breakstone:1995ug+Drijard:1982zk+Angelis:1978ec+Busser:1976su+Akesson:1984iq+Eggert:1975cq+Alper:1975jm+Alper:1974rw} 
that has systematic uncertainties at the same level of precision as the RHIC data.  Even compiling 
a composite spectrum from all available publications results in minimum systematic 
uncertainties of 20-30\%.  Therefore, the significance of $R_{AA}$ measurements at this beam 
energy may remain small until a reference run at RHIC can be performed.  Nevertheless, given these large 
uncertainties, preliminary results presented at the RHIC/AGS Users Meeting in May 2004 show that even at \sqrtsNN = 62.4 
GeV there is substantial suppression for \pT\gt6 GeV/c\cite{RHICAGSUsersSTAR,RHICAGSUsersPHENIX,RHICAGSUsersPHOBOS,Back:2004ra}.  Figure 
\ref{Fig9} shows preliminary results on $R_{AA}$ (left) and $R_{CP}$ (right) for $h^{\pm}$ measured by the STAR experiment at midrapidity 
($|\eta|<0.5$) and slightly forward ($0.9<|\eta|<0.5$)\cite{RHICAGSUsersSTAR}.  The intermediate \pT\ region is less suppressed at 
\sqrtsNN=62.4 GeV than 200 GeV, but for \pT\gt 6 GeV/c, the suppression is comparable and may also be approximately 
\pT-independent.  

This beam energy and \pT-independence of $R_{AA}$ is intriguing.  One possibility is that it may be due simply to the 
interplay of the evolving power-law shape of the underlying partonic spectrum and the medium-induced energy loss for media 
of varying size and density\cite{Eskola:2004cr}.  For central Au+Au at \sqrtsNN=200 GeV, calculations with a 
time-averaged transport coefficient of $\hat{q}$=5-15 GeV$^2$/fm for the medium and a pQCD parton spectrum from PYTHIA 
reproduce the observed $R_{AA}$ for \pT\gt5 GeV/c and are \pT-independent even to the highest accessible \pT.  This has 
intriguing consequences for $R_{AA}$ at the LHC:  the strong suppression in the denser, more opaque system expected in 
Pb+Pb collisions at \sqrtsNN = 5.5 TeV is balanced by the shallower partonic spectrum, which results in more high \pT\ hadrons from 
hard-scatterings near the surface.  The overall effect is $R_{AA}$ with similar shape and magnitude at RHIC and LHC, in contrast to 
other models\cite{Vitev:2002pf}.  It is suggested that $R_{AA}$ may only be able to set a lower limit for the system density when 
$\hat{q}\geq$ 4 GeV$^2$/fm\cite{Eskola:2004cr}.

\begin{figure}
\centering
\includegraphics[width=.45\textwidth]{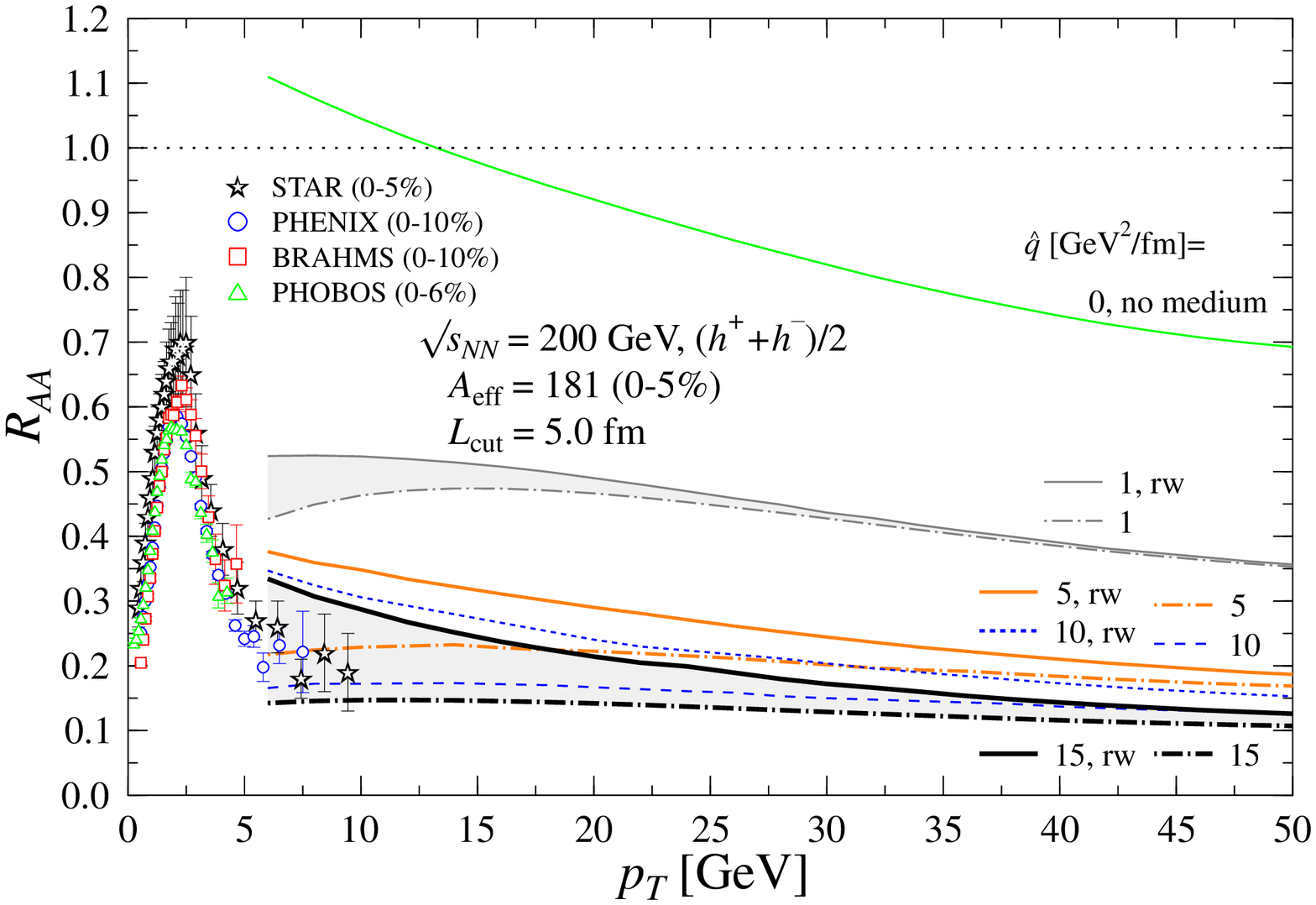}
\vspace{2mm}
\includegraphics[width=.45\textwidth]{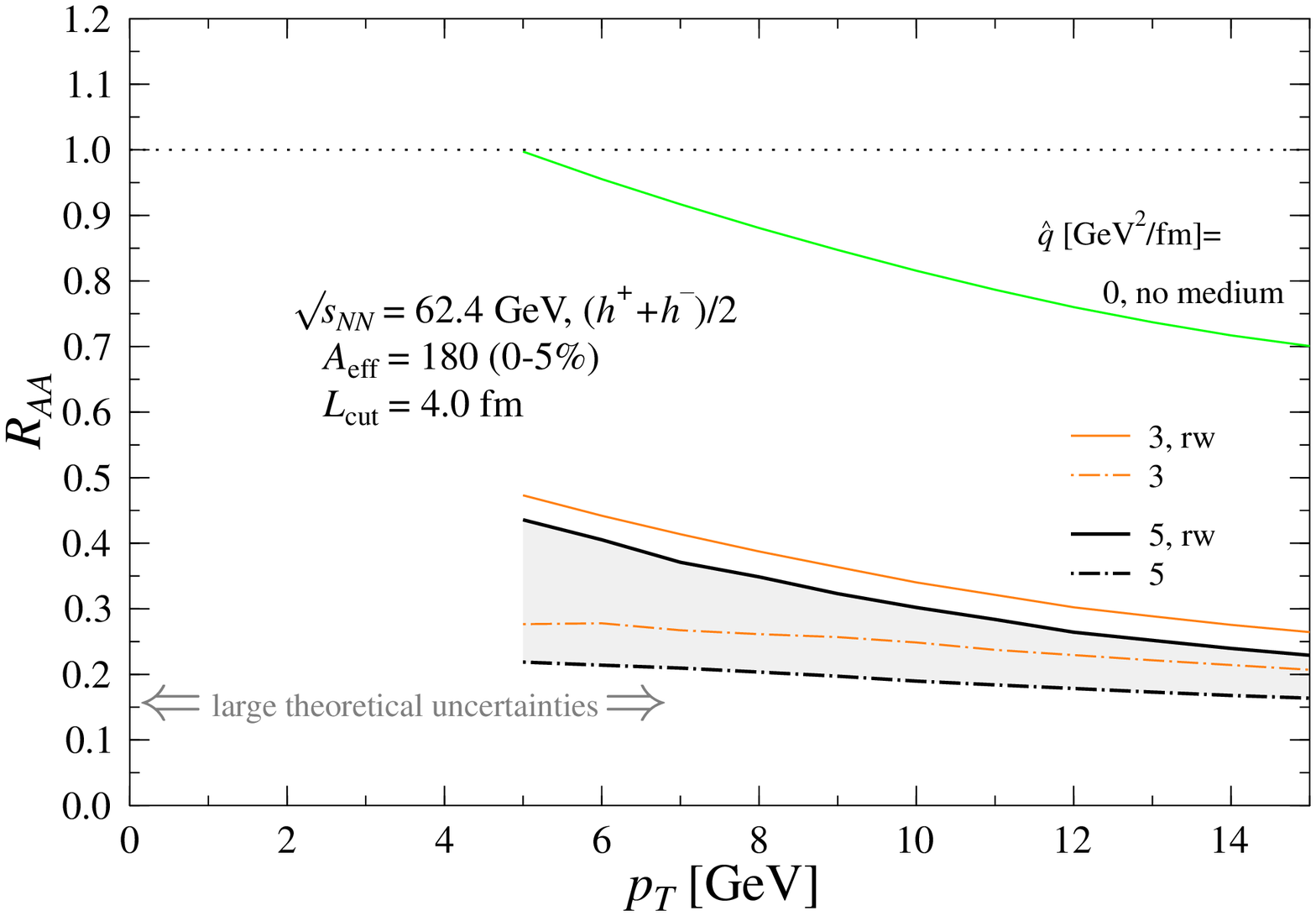}
\caption{\label{Fig10} The \pT\ dependence of $R_{AA}$ at high \pT\ compared to RHIC data at \sqrtsNN = 200 
GeV (left) and predictions for $R_{AA}$ at \sqrtsNN = 62.4 GeV within the same model from Ref.\cite{Eskola:2004cr} (right).}
\end{figure}

At \sqrtsNN = 62.4 GeV, the transport coefficient is probably reduced somewhat and may offer more sensitivity to the medium 
properties, but again it competes with the steeper partonic spectrum.  Figure \ref{Fig10} shows $R_{AA}$ expected for 
\sqrtsNN = 62.4 GeV Au+Au assuming $\hat{q}\simeq$ 3-5 GeV$^2$/fm\cite{Eskola:2004cr}.  The suppression is substantial, 
\pT-independent and reasonably consistent with the preliminary data for \pT\gt6 GeV/c and with several other recently published 
predictions\cite{Vitev:2004gn+Adil:2004cn+Wang:2004yv+Dainese:2004te}.  Clearly, the body of data collected at RHIC over many 
systems and beam energies provides important feedback for testing and refining these models so that reliable estimates of the 
medium properties may be extracted from them.

\section{Conclusion}
Systematic high-precision studies of single hadron inclusive spectra measured in p+p, d+Au and Au+Au collisions at RHIC have yielded several exciting observations:  
the \pT-independent factor 4-5 suppression for \pT\gt6 GeV/c, the complex 
composition of hadron species at intermediate 2\lt\pT\lt 6 GeV/c and the possible beam energy independence of 
$R_{AA}$(\pT\gt6GeV/c) for 62.4\lt\sqrtsNN\lt200 GeV.   These features have stimulated extensive theoretical feedback to 
explain their origins.  The body of data supports the idea that strong final-state partonic interactions are at work in the 
dense medium created in these collisions.   In order to take this qualitative statement closer to quantitative evaluation 
of the medium properties, a comprehensive, coherent explanation of all the high \pT\ phenomena: single inclusive hadron 
spectra, di-hadron correlations/modifications and azimuthal anisotropic emission at high \pT\ must be developed.  As the 
simplest and most investigated of the observables, the single hadron spectra provide a strong baseline for these more detailed and differential
explorations of the medium.  The quality and quantity of the spectra will grow and improve as the RHIC program proceeds, transforming them from the pool of
discovery they provided in the first four years of RHIC running to a tool of precision for future high \pT\ studies in heavy ion collisions at RHIC and beyond.

\ack
This work was performed under the auspices of the U.S. Department of 
Energy by University of California, Lawrence Livermore National Laboratory 
under Contract W-7405-Eng-48.


\end{document}